\documentstyle[aps,pre,epsf,twocolumn,floats,graphicx]{revtex}

\graphicspath{{Figures/}{.}}
\begin{document}
\bibliographystyle{prsty}
\twocolumn[\hsize\textwidth\columnwidth\hsize\csname
@twocolumnfalse\endcsname
\draft

\title{\bf On the Origin of Traveling Pulses in Bistable Systems}

\author{C. Elphick}
\address{Centro de Fisica No Lineal y Sistemas Complejos de Santiago, Casilla
17122, Santiago, Chile}

\author{A. Hagberg\thanks{\tt aric@lanl.gov }}
\address{Center for Nonlinear Studies and T-7, Theoretical Division,
         Los Alamos National Laboratory, Los Alamos, NM 87545}

\author{B. A. Malomed\thanks{\tt malomed@eng.tau.ac.il}}
\address{Department of Interdisciplinary Studies, Faculty of Engineering,
Tel Aviv University, Ramat Aviv 69978, Israel}

\author{E. Meron\thanks{\tt ehud@bgumail.bgu.ac.il}}
\address{The Jacob Blaustein Institute for Desert Research and
         the Physics Department, Ben-Gurion University, \\
         Sede Boker Campus 84990, Israel}

\date{\today}

\maketitle

\begin{abstract}
The interaction between a pair of Bloch fronts forming a traveling domain in a
bistable medium is studied. A parameter range beyond the nonequilibrium
Ising-Bloch bifurcation is found where traveling domains collapse. Only
beyond a second threshold the repulsive front interactions become strong
enough to balance attractive interactions and asymmetries in front speeds, and
form stable traveling pulses. The analysis is carried out for the forced
complex Ginzburg-Landau equation. Similar qualitative behavior is found in the
bistable FitzHugh-Nagumo model.
\end{abstract}

\pacs{PACS number(s):}

\vskip2pc]
\narrowtext

%
%

Traveling waves far from equilibrium are often formed 
when a uniform state is destabilized by a Hopf bifurcation occurring at a 
finite wavenumber~\cite{CH}. Traveling waves or pulses also form 
from  parity breaking bifurcations of stationary patterns~\cite{CI}. 
A related mechanism
that has not received adequate attention involves a parity breaking {\em front
bifurcation} in which a stationary front solution loses stability to a pair of
counter-propagating front solutions~\cite{IMN,CLHL,HM1,BRSP}. 
This bifurcation, sometimes referred to as
the nonequilibrium Ising-Bloch (NIB) bifurcation, has been found in chemical
reactions~\cite{HBKR,Haim} and in liquid crystals~\cite{FRCG,NYK}. 
Bistable systems, which do not necessarily support stationary 
patterns, may exhibit traveling pulses and waves beyond the NIB bifurcation.
Activator-inhibitor systems with non-diffusing
inhibitors provide a good example. For fast inhibitor kinetics 
initial domain patterns always coarse grain and converge to a 
uniform state. For sufficiently slow kinetics, and beyond the NIB bifurcation,
traveling pulses, periodic wavetrains, and spiral waves appear.

Numerical studies of systems with a NIB bifurcation indicate that 
traveling pulses do not appear immediately at the front bifurcation point. 
Instead, there is
an intermediate parameter range where initial domains may travel
but eventually collapse. Only past a second threshold parameter value
do initial domains converge to stable traveling pulses~\cite{HM1}. 
In this paper we study the interactions between a pair of traveling 
fronts in this intermediate parameter range. 
We find that the balance of repulsive front interactions with 
attractive interactions and an asymmetry between leading and 
trailing fronts gives this threshold parameter value.

We choose to study the parametrically forced complex Ginzburg-Landau 
(CGL) equation 
\begin{eqnarray}
\label{cgl} 
A_t&=& (\mu+i\nu)A+(1+ic_1)A_{xx}-(1+ic_3)|A|^2A \\
   & & \mbox{} +\gamma A^{*}+\alpha\,\nonumber,
\end{eqnarray}
where $A(x,t)$ is a complex field and 
$\nu,c_1,c_3$ and $\gamma$  are real parameters. 
The parameter $\alpha$ can be a complex 
number, but since the final results we present here do not depend on its 
imaginary part we assume $\alpha$ is also real~\cite{com0}. 
The CGL equation ($\gamma=\alpha=0$) is often obtained as an envelope 
equation for an extended system undergoing a Hopf bifurcation at zero
wavenumber~\cite{Ku}. Then, the variable  $A(x,t)$ describes 
weak modulations of the homogeneous oscillations. The terms $\alpha$ and
$\gamma A^{*}$  in~(\ref{cgl}) represent, respectively, the effect of 
parametric forcing with equal and twice the
system's natural oscillation frequency~\cite{EIT,com1}. 
Equation~(\ref{cgl}) has been introduced recently in the
context of liquid crystals~\cite{FG}. 

The parametric forcing term $\gamma A^{*}$ breaks the phase
shift symmetry, $A\to Ae^{i\phi}$, of Eqn.~(\ref{cgl}) and 
reduces the one-parameter family of cw solutions of the CGL equation, 
$A=A_0e^{i(\nu- c_3\mu)t+i\phi}$, $0<\phi<2\pi$, 
to two pairs of stable-unstable solutions with fixed $\phi$ 
values, arising in saddle-node bifurcations. Equation~(\ref{cgl}) 
therefore describes a bistable extended system of
two stable uniform states that oscillate with different phases. 
The second forcing term, $\alpha$, breaks the
parity symmetry ($A\to -A$) of these two states. 
The front solutions we will
be concerned with connect these two states at $x\to\pm\infty$.

A simpler, gradient version of Eqn.~(\ref{cgl}) is obtained by omitting the 
linear and nonlinear dispersion terms:
\begin{equation}
A_t\,=\,\mu A+A_{xx}-|A|^2A+\gamma A^{*}+\alpha\,.
\label{gl}
\end{equation}
A physical application of~(\ref{gl}) is Rayleigh-B\'{e}nard 
convection with periodic spatial modulation of the cell height~\cite{KP}
or heating~\cite{LG}.
When $\alpha=0$, Eqn.~(\ref{gl}) has three types of stationary
front solutions
\begin{eqnarray}
I(x;\sigma)&=&\sigma A_0\tanh\big({1\over\sqrt{2}}A_0 x\big)
\,, \label{Is}\\
B_\pm(x;\sigma)&=&\sigma A_0\tanh(kx)
\pm i\sqrt{\mu-3\gamma}~{\rm sech}(kx)\,, \label{Bl}
\end{eqnarray}
where $A_0=\sqrt{\mu+\gamma}$, $k=\sqrt{2\gamma}$  and  $\sigma=\pm 1$ is 
the front polarity which stems from the reflection 
symmetry $x\to -x$ of Eqns.~(\ref{gl}) and~(\ref{cgl}). The front solutions
$I(x;\sigma)$ and $B_\pm(x;\sigma)$ are
equivalent to N\'eel (Ising) and Bloch domain walls in ferromagnets with weak
anisotropy~\cite{CLHL} and will be referred to as Ising and Bloch fronts.
The Ising front $I(x;\sigma)$ loses stability as $\gamma$ is 
decreased past the critical value $\gamma_c=\mu/3$. 
At that point the two Bloch
fronts $B_\pm(x;\sigma)$ appear and are stable~\cite{CLHL}. 

\begin{figure}
\includegraphics[width=3.5in]{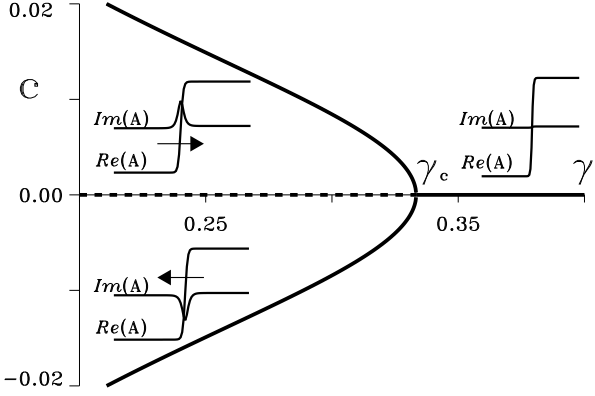}
\caption{The nonequilibrium Ising-Bloch (NIB) bifurcation for front solutions
of Eqn.~(\protect\ref{cgl}).  For $\gamma>\gamma_c$ there is a single
stable Ising front with zero speed (solid line).  
For $\gamma<\gamma_c$ the Ising front is unstable (dashed line) 
and a pair of stable counterpropagating Bloch fronts appears (solid lines).
Parameters: $\mu=1.0$, $\nu=0.01$, $c_1=c_3=\alpha=0.0$.
}  
\label{fig:pfork}
\end{figure}

The non-gradient terms associated with $\nu, c_1$
and $c_3$ remove the degeneracy of the three stationary 
solutions~(\ref{Is}) and~(\ref{Bl}). 
With any of these terms nonzero,
the two Bloch fronts propagate in opposite directions 
at a speed proportional to the corresponding
coefficient, $\nu, c_1$ or $c_3$~\cite{CLHL}. 
In that case, a plot of the front velocity, $c$, versus $\gamma$ 
yields the NIB bifurcation diagram
shown in Fig.~\ref{fig:pfork}.

To study front interactions we consider the symmetric 
($\alpha=0$) and nearly gradient case, where front solutions 
of~(\ref{cgl}) can be expanded around front 
solutions of the gradient system~(\ref{gl}). 
We introduce a small parameter $\epsilon\ll 1$ and  assume that the constants 
$\nu, \alpha, c_1$, and $c_3$ are all of order $\epsilon$. We also assume 
proximity to the Ising-Bloch bifurcation point, $\mu-3\gamma\sim\sqrt\epsilon$.
A traveling domain solution of Eqn.~(\ref{cgl}) is sought as
\begin{eqnarray}
\label{ansatz}
A(x,t)&=&B_+[x-x_l(T);+1] + B_-[x-x_r(T);-1]\\ 
&&\mbox{}- A_0 + R(x,T)\,\nonumber, 
\end{eqnarray}
where $x_r$ and $x_l$  
are the positions of the leading (right) and trailing
(left) Bloch fronts, $T=\epsilon t$ is a slow time, $B_\pm$ are given 
by~(\ref{Bl}), and $R$ is a small correction term of order $\epsilon$. 
The two polarities ($\sigma=\pm 1$)
are necessary to construct a {\em domain} bounded by the fronts. The two 
types of Bloch fronts, $B_-$ and $B_+$, 
make the domain {\em traveling} instead of shrinking or 
expanding. We assume that the domain is much wider than the width 
of the fronts, $k^{-1}$, or more accurately, that 
$\exp[-2k(x_r-x_l)]\sim\epsilon\ll 1$.

Our objective is to derive equations of motion for the front positions, $x_l$
and $x_r$. Using the anstaz~(\ref{ansatz}) in~(\ref{cgl}) and following the 
method of Refs.~\cite{EMS}, we find the following solvability condition
for the right (leading) front:
\begin{eqnarray}
\label{solvability}
&&\dot x_r\int_{-\infty}^\infty\left|{\partial B_-\over\partial x}\right|^2dx
+\alpha \int_{-\infty}^\infty{\partial B_-^*\over\partial x}dx \\
&&\mbox{}+i\nu\int_{-\infty}^\infty{\partial B_-^*\over\partial x}B_-dx
+ic_1\int_{-\infty}^\infty{\partial B_-^*\over\partial
x}{\partial^2B_-\over\partial x^2}dx \nonumber \\
&&\mbox{}-ic_3\int_{-\infty}^\infty{\partial B_-^*\over\partial x}\left| B_-\right|^2B_-dx
+\int_{-\infty}^\infty{\partial B_-^*\over\partial x}{\cal N}dx \nonumber \\
&&\mbox{}+c.c. = 0\,\nonumber,
\end{eqnarray}
where $c.c.$ stands for the complex conjugate. 
A similar condition is obtained for the left (trailing) front. 
In Eqn.~(\ref{solvability}) the dot over $x_r$ represents 
the derivative with respect to the fast time $t$, and
${\cal N}$ is a nonlinear interaction term 
\begin{eqnarray}
{\cal N}&=&(B_-^*+B_+^*)(A_0-B_+)(B_--A_0)\nonumber\\
&\mbox{}+&(B_-+B_+)(A_0-B_+)(B_-^*-A_0)\nonumber \\ 
&\mbox{}+&(B_-+B_+)(A_0-B_+^*)(B_--A_0)\,\nonumber,
\end{eqnarray}
where the arguments of $B_\pm$ are as in~(\ref{ansatz}). Analytical 
evaluation of the
nonlinear interaction integral in~(\ref{solvability}) leads to the following
equations, with $\eta=\sqrt{\mu-3\gamma}$~\cite{EMS}:
\begin{eqnarray}
\label{motion}
{4\over 3}k A_0\dot x_r&=&2\alpha +\pi\eta[-\nu+c_3\mu+(c_1-c_3)\gamma]\\
&&\mbox{}-8A_0^3 e^{-2k(x_r-x_l)}+4A_0\eta^2 e^{-k(x_r-x_l)}\,,\nonumber\\
{4\over 3}k A_0\dot x_l&=&-2\alpha +\pi\eta[-\nu+c_3\mu+(c_1-c_3)\gamma]\nonumber\\
&&\mbox{}+8A_0^3 e^{-2k(x_r-x_l)}-4A_0\eta^2 e^{-k(x_r-x_l)}\,.\nonumber
\end{eqnarray} 

Combining Eqns.~(\ref{motion}) gives a single equation of motion 
for the distance between the two fronts,
$L=x_r-x_l$:
\begin{equation}
k A_0\dot L = 3\alpha -12 A_0^3 e^{-2kL} + 6A_0\eta^2 e^{-kL}\,.
\label{L}
\end{equation}
The first term on the right hand side describes the effect of the broken
symmetry between the two Bloch fronts; the initial domain expands (shrinks) 
in time when the leading front is faster (slower) than the trailing front. 
The second term describes an attractive front interaction generated by the 
real parts of the Bloch front solutions. The last term, generated by the
imaginary parts of the Bloch front solutions, describes a longer range
repulsive interaction. The repulsive interaction strengthens as $\gamma$ is 
decreased below the Ising-Bloch
bifurcation point, $\gamma_c=\mu/3$, and becomes dominant at sufficiently
small $\gamma$ values.

Notice that in Eqn.~(\ref{L}) the non-gradient terms 
associated with $\nu, c_1$, and $c_3$ have disappeared. 
To leading order the effect of these terms is to grant the two Bloch fronts
equal speeds. Therefore the distance between the fronts is not affected.
This suggests that the existence and stability of pulse solutions (moving
domains) of Eqn.~(\ref{cgl}) can be studied starting from the
gradient version~(\ref{gl}). The latter can be written as 
$A_t=-\delta H/\delta A^{*}$, where $\delta/\delta  
A^{*}$ stands for the variational derivative, and the Lyapunov function 
(pseudo-Hamiltonian) is
\begin{eqnarray}
\label{density}
H\,&=&\,\int_{-\infty}^{+\infty}{\cal H}dx\,,\label{H}\\
{\cal H}\,&=&\,|A_x|^2-\mu|A|^2+\frac{1}{2}|A|^4 \nonumber\\
&&\mbox{}-\frac{1}{2}\gamma [A^2+(A^{*})^2]-\alpha(A+A^*)\,.\nonumber
\end{eqnarray}
This gradient representation implies relaxational dynamics toward a minimum
of $H$. The part of $H$ which depends upon the separation 
distance $L$ between the Bloch fronts can be found following Ref.~\cite{we}.
The resulting effective {\it pseudopotential} of the interaction for
$\mu\approx 3\gamma$ is
\begin{equation}
U(L)=
\frac{4A_0^2}{3\sqrt{3}}
\left(-{4}\sqrt{2}e^{-2kL}+
\frac{2\sqrt{2}\eta^2}{k^2}e^{-kL}
-\frac{\alpha}{k^2}L\right)\,. 
\label{UL}
\end{equation}
The extrema of $U(L)$ coincide with stationary solutions of~(\ref{L}).
%
%

\begin{figure}
\includegraphics[width=3.5in]{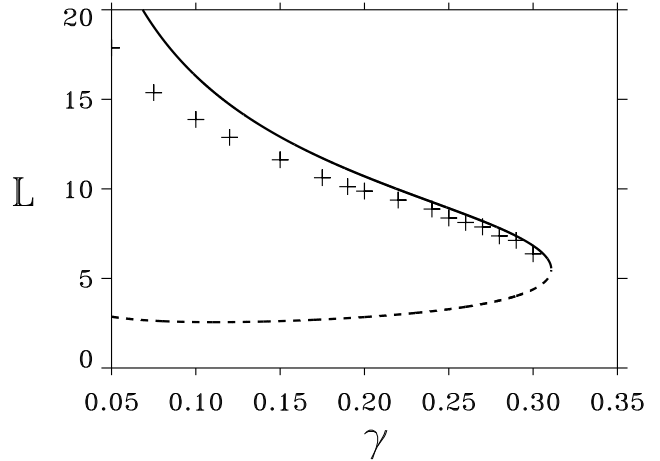}
\vspace{0.1cm}
\caption{The distance, $L$, between the front and back of a pulse
solution for $\gamma$ near the NIB bifurcation.  The solid and dashed lines
represent the stable and unstable branches solutions from
Eqn.~(\protect\ref{Lsol}).  The crosses are data from direct numerical
solution of Eqn.~\protect(\ref{cgl}).  
Parameters: $\mu=1$, $\nu=0.01$, $\alpha=-0.001$, $c_1=c_3=0$.
}
\label{fig:Lgamma}
\end{figure}

\begin{figure}
\includegraphics[width=3.0in]{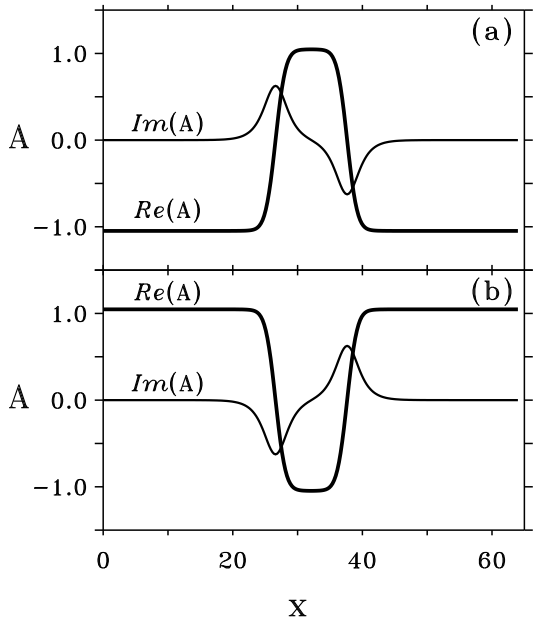}
\caption{Pulse solutions to Eqn.~(\protect\ref{cgl}) near the NIB bifurcation.
(a) A stable ``up'' pulse, $\alpha < 0$. (b) A stable ``down'' pulse 
$\alpha > 0$.}
\label{fig:shape}
\end{figure}

To find  traveling pulse solutions of~(\ref{cgl}), which correspond to  
stationary (fixed point) solutions of~(\ref{L}), 
we set $\dot L=0$ in ~(\ref{L}) or $dU/dL=0$ in
~(\ref{UL}) and solve the 
resulting quadratic equation for $z=\exp(-kL)$. The solutions are:
\begin{equation}
L=-k^{-1}\ln\big(\eta^2\pm\sqrt{\eta^4+4A_0\alpha}\bigr)+2k^{-1}\ln{2A_0}\,.
\label{Lsol}
\end{equation}
In the symmetric case, $\alpha=0$, and for $\eta>0$, there is only one (finite)
pulse solution. A linear stability analysis indicates the solution is 
unstable. This result complies with an earlier finding
reported in Ref.~\cite{KRT}.
For $\alpha>0$ the leading front is faster than the trailing front. Again, 
only one pulse solution exists and a linear stability analysis indicates 
the solution is unstable. The same conclusion follows from a graph of Eq. 
(\ref{UL}): the single pulse solution corresponds to a maximum of the 
interaction pseudopotential $U(L)$.  
For $\alpha<0$ no pulse solutions exist if
$\eta^4<4A_0\vert\alpha\vert$. A saddle-node pair of pulse solutions appears at
$\gamma=\gamma_p(\alpha)$, where $\gamma_p(\alpha)$ solves
$(\mu-3\gamma_p)^2=4\sqrt{\mu+\gamma_p}\vert\alpha\vert$. Graphs of these 
solutions in the $L-\gamma$ plane are shown in Fig.~\ref{fig:Lgamma}.
The upper and lower branches represent stable and unstable solutions,
and pertain to a minimum and a maximum of $U(L)$, respectively. Also shown 
in Fig.~\ref{fig:Lgamma} are results from direct
numerical solutions of Eqn.~(\ref{cgl}) 
showing the stable pulse branch. 
The agreement for small $\eta$ is within 5\%. The shape of the stable traveling
pulse is shown in Fig.~\ref{fig:shape}a.

\begin{figure}
\includegraphics[width=3.5in]{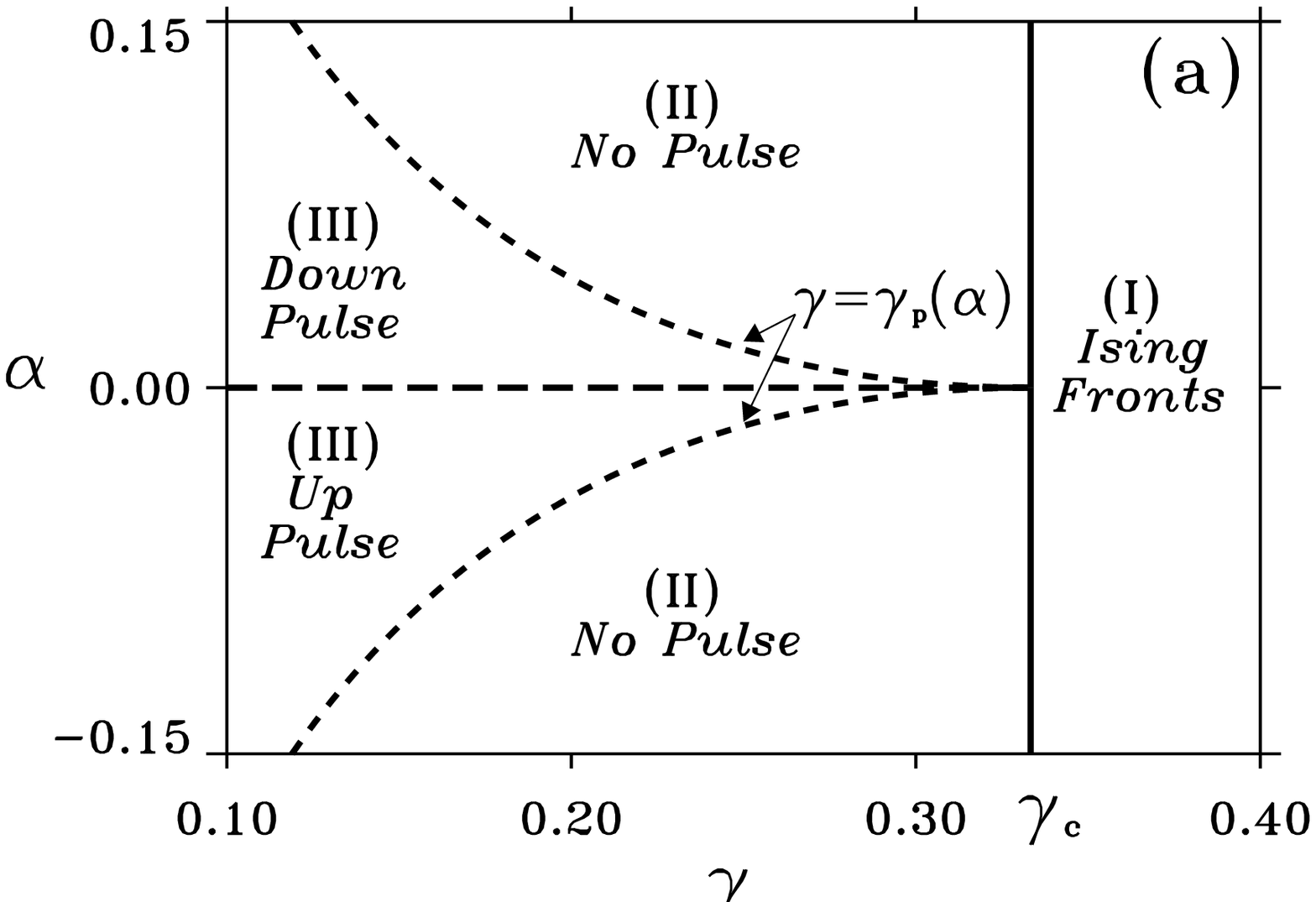}
\includegraphics[width=3.5in]{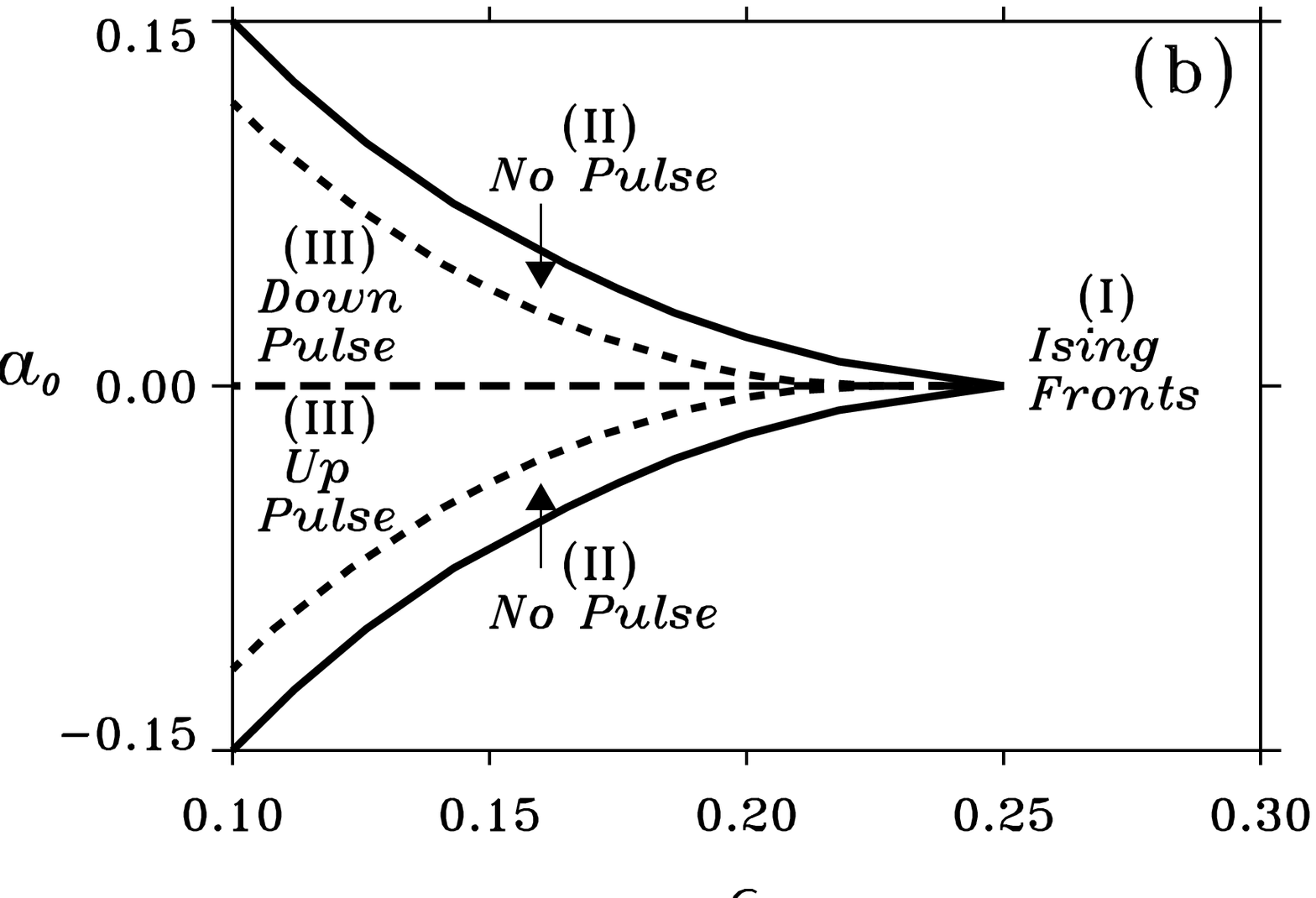}
\vspace{0.2cm}
\caption{(a)
A phase diagram in the plane of the forcing parameters,
$\alpha-\gamma$, showing the regions of stable front and  pulse solutions
to the forced CGL equation.  
For $\gamma>\gamma_c$ only Ising fronts exist and 
there are no pulse solutions.  For $\gamma<\gamma_c$
a pair of stable Bloch fronts exist but they only form a stable traveling pulse
solution when $\gamma<\gamma_p$. Parameters: $\mu=1.0$, $\nu=0.05$, $c_1=c_3=0$.
(b) The bistable FHN equation (\protect\ref{FHN}) shows the same
type of phase diagram in the $\epsilon-a_0$ parameter plane.  
Parameters: $a_1=2.0$, $\delta=0$.
}
\label{fig:cgl-fhn}
\end{figure}

The conclusion that no stable pulses exist for $\alpha>0$ is a result of the
specific ansatz~(\ref{ansatz}) for an ``up'' pulse as shown 
Fig.~\ref{fig:shape}a.
From the symmetry $A\to -A$ of~(\ref{cgl}), a symmetric stable ``down'' pulse
exists for $\alpha>0$. The shape of this pulse is displayed in
Fig.~\ref{fig:shape}b. 
A phase diagram of all stable front and pulse solutions is shown in
Fig.~\ref{fig:cgl-fhn}a. 
Three main regions can be distinguished: (I) The entire Ising regime, 
$\gamma>\gamma_c=\mu/3$, where only a stable Ising front solution exists. 
In this region domains shrink or expand but do not travel or form pulses. 
(II) A region in the Bloch regime, $\gamma_p(\alpha)<\gamma<\gamma_c$,
where a pair of stable counterpropagating Bloch fronts exist.
In this region domains travel but do not form pulses; they either expand to
infinity or shrink and collapse.
(III) The rest of the Bloch regime, $\gamma<\gamma_p(\alpha)$, excluding the
$\alpha=0$ line (the $\gamma$ axis), where stable
traveling pulses exist in addition to the pair of Bloch fronts. 
The $\gamma$ axis separates regions of up and down pulses.

This behavior is rather general. Shown in
Fig.~\ref{fig:cgl-fhn}b is an analogous phase diagram for the bistable 
FitzHugh-Nagumo (FHN) model
\begin{eqnarray}
u_t&=&u-u^3-v+u_{xx}\,,\nonumber\\
v_t&=&\epsilon(u-a_1v-a_0)+\delta v_{xx}\,,
\label{FHN}
\end{eqnarray}
where $u$ and $v$ are real fields and $\epsilon,\delta,a_1$ and $a_0$ are real
constants. The parameters $\epsilon$ and $a_0$ play a similar role as
$\gamma$ and $\alpha$, respectively, and the same three regions (I), (II) and 
(III) appear in the $\epsilon$-$a_0$ plane.

The existence of an intermediate parameter region where traveling domains
collapse rather than converge to stable pulses, has been observed in numerical
simulations of model equations~\cite{HM1,EiEr}. 
It is also expected to be observed in a number of 
experimental systems including catalytic reactions on platinum 
surfaces~\cite{EiEr}, liquid crystals 
subjected to rotating magnetic fields~\cite{FRCG,NYK,MiMe}, 
oscillatory chemical reactions subjected to periodic forcing~\cite{Valery}, and
the Ferrocyanide-Iodate-Sulfite reaction~\cite{LeSw}.


\begin{references}
\bibitem{com0} The imaginary part of $\alpha$ may contribute, however, to 
higher order corrections not considered considered here.
\bibitem{com1} Equation (\ref{cgl}) is obtained assuming that the detuning 
$\nu$
of the forcing at the system's natural frequency is exactly half the detuning
of the forcing at twice the natural frequency.
\end{references}
\end{document}